\documentclass[aip,amsmath,amssymb,preprint]{revtex4-1}
\usepackage{graphicx}
\graphicspath{{Figures/}}
\usepackage{dcolumn}
\usepackage{bm}
\usepackage[utf8]{inputenc}
\usepackage[T1]{fontenc}
\usepackage{mathptmx}
\usepackage{etoolbox}
\usepackage{subcaption}
\usepackage{float}
\usepackage{booktabs}
\usepackage[hidelinks]{hyperref}

\makeatletter
\def\@email#1#2{%
 \endgroup
 \patchcmd{\titleblock@produce}
  {\frontmatter@RRAPformat}
  {\frontmatter@RRAPformat{\produce@RRAP{*#1\href{mailto:#2}{#2}}}\frontmatter@RRAPformat}
  {}{}
}
\makeatother

\begin{document}

\preprint{AIP/123-QED}

\title{Numerical Investigation of Boundary-Layer Height and Actuation-Parameter Effects of a Circular Synthetic Jet Actuator in Crossflow}

\author{Howard Haonan Ho}
\email{howard.ho@mail.utoronto.ca}
\affiliation{Department of Mechanical and Industrial Engineering, University of Toronto, 5 King’s College Road, Toronto, Ontario M5S 3G8, Canada}

\author{Ebenezer Ekow Essel}
\affiliation{Department of Mechanical, Industrial and Aerospace Engineering, Concordia University, 1515 St. Catherine W., Montreal, Quebec H3G 1M8, Canada}

\author{Pierre Edward Sullivan}
\affiliation{Department of Mechanical and Industrial Engineering, University of Toronto, 5 King’s College Road, Toronto, Ontario M5S 3G8, Canada}
\altaffiliation{ASME Fellow}

\date{\today}

\begin{abstract}
Three-dimensional unsteady numerical simulations are performed to investigate the effects of blowing ratio $C_B$ ($0.85 < \overline{U}_j/U_\infty < 1.7$), stroke ratio $L^+$ ($10.6 < \overline{U}_j /(fd) < 21.3$), and boundary-layer height ratio $D^+$ ($2.1<\delta/d<8.0$) on circular synthetic jet actuator (SJA) performance in crossflow. Nine cases are examined at constant free-stream velocity $U_\infty$, with systematic independent variation of averaged jet velocity $\overline{U}_j$, actuation frequency $f$ ($200$-$400~\mathrm{Hz}$), and boundary-layer momentum thickness Reynolds number ($170<Re_\theta<740$) to isolate the individual effects of these parameters on a circular-nozzle SJA with fixed nozzle diameter $d$ in crossflow. Instantaneous vortical structures exhibited tilted vortex rings with a trailing vortex pair at low actuation frequency; closely packed expelled vortical structures for higher frequency SJAs, and the largest boundary-layer height ratio induced hairpin-like vortices. Near-wall tertiary vortices, which promote downwash and increase wall shear stress, remain coherent longer and have extended spanwise coverage for low $D^+$. Time-averaged boundary-layer profiles and skin-friction distributions reveal that SJAs with low to moderate $D^+$ have the greatest potential for separation control, maintaining increased near-wall momentum over extended streamwise distances.
\end{abstract}

\maketitle

\section{\label{sec:intro}Introduction}

Synthetic jet actuators (SJAs) are zero-net-mass-flux (ZNMF) devices that have gained significant attention in active flow control due to their compactness, energy efficiency, and ability to operate without additional fluid supply systems \cite{amitay1998,smith1998,xia2017,glezer2002}. Unlike conventional steady jets that require continuous input from pressurized reservoirs or plumbing, SJAs operate with the surrounding bulk fluid, making them highly adaptable for embedded aerodynamic control applications. By transferring linear momentum into the flow via the periodic expulsion of vortex structures, SJAs have demonstrated effectiveness in diverse control scenarios such as separation control \cite{devanna2025,yen2012,tadjfar2020,machado2024,jia2024}, fluid mixing enhancement \cite{zhang2021,he2023,huang2019,huang2016,huang2013}, and turbulence manipulation \cite{palumbo2022}. Their ability to generate significant momentum flux while consuming minimal power makes them promising for low-Reynolds-number applications in UAVs \cite{jafari2023,xu2025,nakamura2025}, wind turbines \cite{maldonado2009,aguirre2025,saemian2025}, and compact heat exchangers \cite{arshad2020,lyu2025,hammond2025,dutta2022}.

\begin{figure}[ht]
    \includegraphics[width=0.5\textwidth]{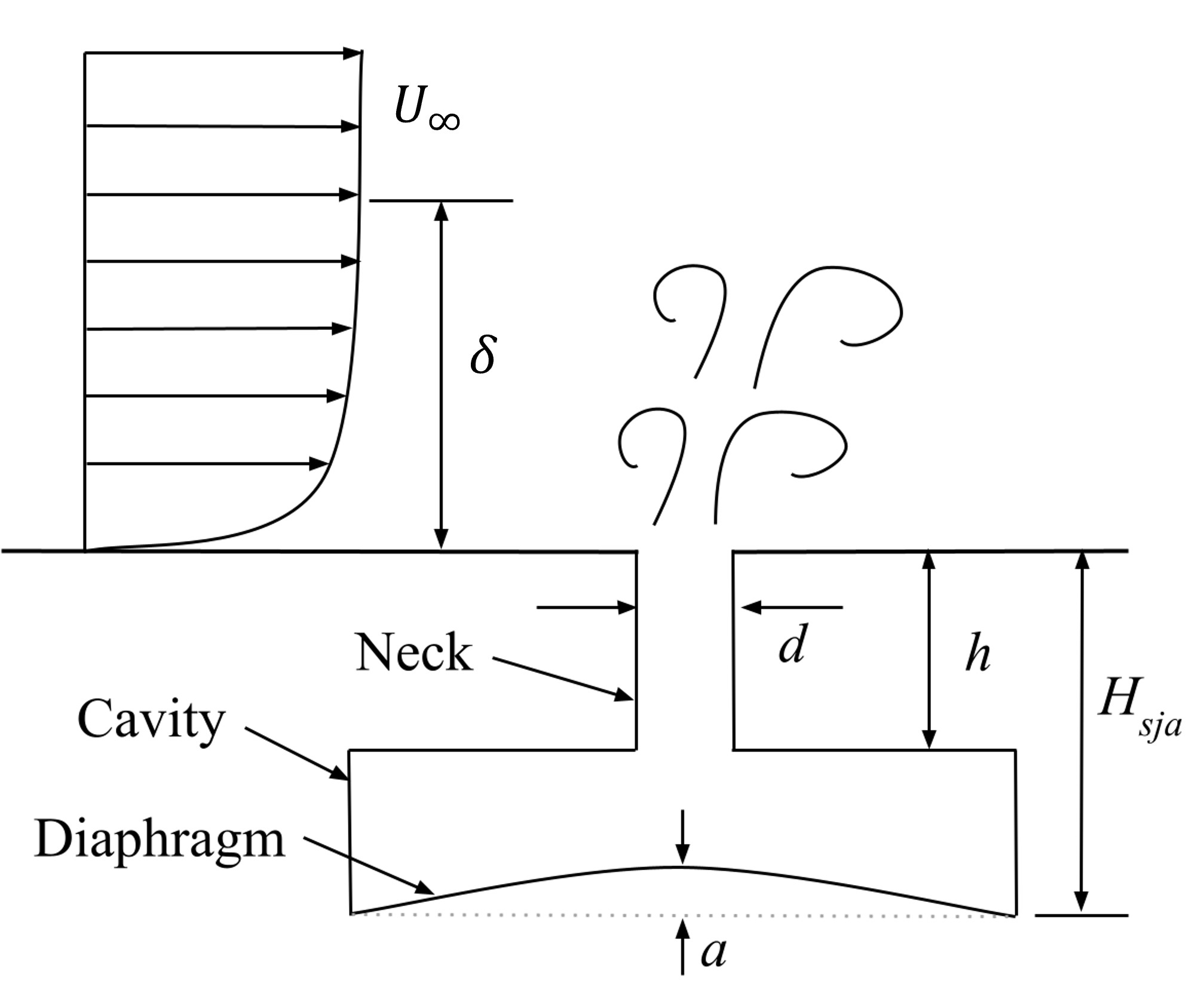}
    \caption{\label{fig:sja_schematic}Schematic of a synthetic jet actuator in crossflow during the expulsion cycle.}
\end{figure}

An SJA typically consists of a cavity with a vibrating diaphragm or piston and a neck leading to an orifice or slot exit on the control surface \cite{cattafesta2011}. A schematic of an SJA in a crossflow boundary layer is shown in Fig.~\ref{fig:sja_schematic}; the crossflow has a free-stream velocity $U_\infty$ and boundary-layer thickness $\delta$. The actuator is characterized by an orifice diameter $d$, neck height $h$, and an overall actuator height denoted here as $H_{\mathrm{sja}}$. During each actuation cycle, the diaphragm alternates between ingestion and expulsion. When the expelled jet column escapes reingestion, a vortex ring forms and constitutes a synthetic jet (SJ). While there is no net mass addition to the flow, each expulsion transfers linear momentum to the external flow through the expelled vortical structures.

\begin{figure}[ht]
    \includegraphics[width=1.0\textwidth]{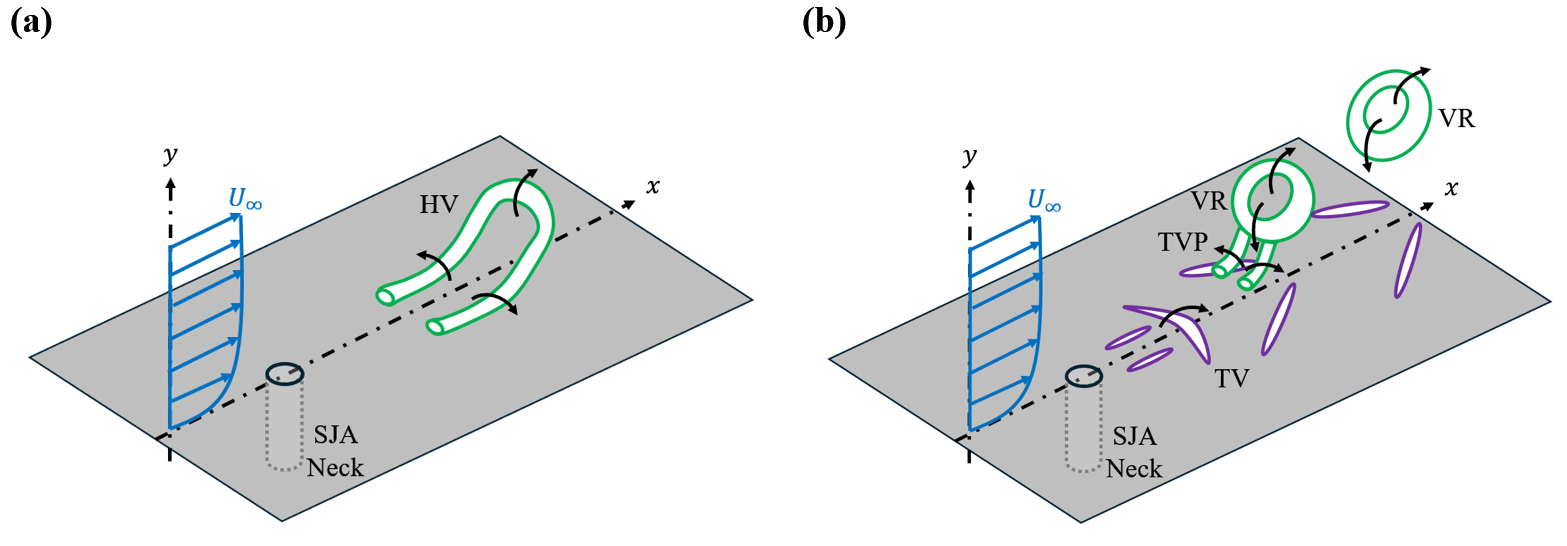}
    \caption{\label{fig:sja_structures_schematic}Schematic of vortex structures formed by a synthetic jet actuator in boundary-layer crossflow: (a) hairpin vortex (HV), or (b) tilted vortex ring (VR) with trailing vortex pair (TVP) and near-wall tertiary vortices (TV).}
\end{figure}

The interaction between synthetic jets and crossflow involves three-dimensional mechanisms of entrainment, penetration, and shear-layer instabilities, generating coherent structures such as upstream horseshoe vortices, hairpins, and vortex rings \cite{jabbal2006,jabbal2008,zhong2005,crook2001}. Depending on operating parameters, the jet column evolves into either a hairpin vortex or a tilted vortex ring with a trailing vortex pair and near-wall tertiary vortices, as shown in Fig.~\ref{fig:sja_structures_schematic}. Key dimensionless parameters governing strength and penetration of the synthetic jet in a crossflow are the blowing ratio $C_B = \overline{U}_j / U_\infty$, stroke ratio $L^+=\overline{U}_j /(fd)$, and momentum coefficient 
\begin{equation}
    C_\mu=\frac{\rho_j \overline{U}_j ^2}{\rho_\infty U_\infty^2}\frac{d}{\theta_0}
\end{equation}
where $\overline{U}_j$ is the average jet velocity during the expulsion cycle, defined as
\begin{equation}
    \overline{U}_j = \frac{2}{T A_n} \int_{A_n}\!\!~\int_0^{T/2} U_j(t, A_n)\, dt \, dA_n
\end{equation}
$A_n$ is the jet exit area, $T$ the actuation period, $f$ the actuation frequency, and $\theta_0$ the baseline momentum thickness. Prior experimental and numerical studies show strong sensitivity of structures to these parameters \cite{crook2001,tang2005,zhong2005a}.

Jabbal and Zhong \cite{jabbal2008} used dye visualization to reveal flow regimes under different ~$C_B$. At low $C_B$ ($<0.35$), expelled vortices form stretched hairpins that convect near the wall \cite{zhong2005}. With increasing $C_B$, tilted vortex rings form and penetrate further \cite{jabbal2008}, often accompanied by trailing vortex pairs and near-wall tertiary vortices that promote downwash \cite{zhou2008,jabbal2010}. Ho et al. \cite{ho2022} used 3-D URANS to examine the influence of $C_B$ ($0.32$–$1.10$) in turbulent boundary-layer crossflow and observed improved penetration at constant frequency, but a trade-off between mid-span wall-shear increase and spanwise control authority. Actuation frequency is also central to control effectiveness \cite{greenblatt2000,amitay2012}. Excessive frequency can reduce separation control due to insufficient momentum per stroke \cite{zhang2015,kim2022}. In addition to $C_B$, $L^+$ helps determine whether structures appear as stretched hairpins or well-defined vortex rings \cite{jabbal2008}. The effect of $L^+$ ($12$–$42$) at constant $C_B=5$ was studied numerically in \cite{ho2024}, showing reduced penetration and changing $C_f$ with increasing frequency. Boundary-layer characteristics further influence performance. Chaudhry and Zhong \cite{chaudhry2014} examined laminar and turbulent layers with the same $D^+=\delta/d=6$ and observed hairpins, stretched rings, and tilted vortices in both. Higher $C_B$ and $L^+$ produced structures that persisted longer in the turbulent layer.

Table~\ref{tab:literature_review} summarizes previous experimental and numerical studies investigating circular synthetic jet actuators in boundary layer crossflow. The table highlights that while there is substantial prior work examining SJAs at low to moderate blowing ratios, limited studies have explored higher momentum configurations ($C_B > 1$) with extended stroke ratios ($L^+ > 10$). Furthermore, most investigations maintain a fixed boundary layer height ratio, leaving the independent influence of $D^+$ on SJA performance inadequately characterized. This gap poses a direct challenge for applying SJAs in flow control applications, where a small $D^+$ requires a larger array of SJAs to achieve the same spanwise control authority, while a larger $D^+$ can lead to significant alterations to the surface geometry.

\begin{table}[ht]
\caption{\label{tab:literature_review}Summary of previous studies on circular-nozzle SJAs in boundary-layer crossflow. Technique abbreviations: Dye Vis. = Dye Visualization, LDV = Laser Doppler Velocimetry, PIV = Particle Image Velocimetry, HWA = Hot-Wire Anemometry, DNS = Direct Numerical Simulation, LES = Large-Eddy Simulation, URANS = Unsteady Reynolds-Averaged Navier–Stokes.}
\begin{ruledtabular}
\begin{tabular}{lcccccl}
Reference & $Re_\theta$ & $D^+$ & $C_B$ & $L^+$ & Technique \\
\hline
Zhong et al. \cite{zhong2005} & 395--547 & 2.7--3.7 & 0.06--0.7 & 0.56--1.4 & Dye Vis.\\
Shuster et al. \cite{shuster2005} & -- & -- & 1.12 & 1--2 & PIV\\
Schaeffler et al. \cite{schaeffler2006} & -- & 3.3 & 0.56--1.3 & -- & LDV, 2D PIV \\
Dandois et al. \cite{dandois2006} & 4300 & 3 & 1.45 & 17 & URANS, LES \\
Wu \& Leschziner \cite{wu2008} & 920 & 4 & 2 & 91 & LES\\
Jabbal \& Zhong \cite{jabbal2008} & -- & 2.3--3.6 & 0.08--0.7 & 0.8--5.1 & Dye Vis.\\
Wu \& Leschziner \cite{wu2009} & 2400 & 10 & 2 & 91 & LES\\
Jabbal \& Zhong \cite{jabbal2010} & -- & 4 & 0.27--0.54 & 1.6--2.7 & 2D PIV \\
Chaudhry \& Zhong \cite{chaudhry2013} & 320 & 6 & 0.17--0.54& 1.7--2.7 & PIV \\
Chaudhry \& Zhong \cite{chaudhry2014} & 320 & 6 & 0.11--0.36 & 2.2--3.6 & Dye Vis. \\
Xia \& Mohseni \cite{xia2017} & 85--144 & 3 & 2.8 -- 8.3 & 2.8--5.7 & HWA, PIV \\
Palumbo et al. \cite{palumbo2020} & 500--550 & 0.23--0.25 & 0.1 & 0.22--0.85 & DNS \\
Palumbo et al. \cite{palumbo2022} & 550 & 0.23--0.25 & 0.05--0.1 & 0.1--0.85 & DNS \\
Ho et al. \cite{ho2022} & 900 & 7.25 & 0.32--1.10 & 3.9 -- 13  & URANS \\
Ho et al. \cite{ho2024} & -- & 1.5 & 4.9 & 11.5 -- 41 & URANS \\
Chhetri et al. \cite{chhetri2025} & 895 & 7.75 & 0.65 & 4.1 & URANS\\
\hline
Current Study & 170--740 & 2.1--8.0 & 0.85--1.7 & 10.6--21.3 & URANS\\
\end{tabular}
\end{ruledtabular}
\end{table}

Therefore, the objective of this study is to investigate the effects of blowing ratio ($0.85<C_B<1.7$) and stroke ratio ($10.6<L^+<21.3$) on SJA performance across different boundary-layer height ratios ($2.1<D^+<8$) using three-dimensional URANS. Because SJAs often operate near the cavity’s Helmholtz resonance, $C_B$ and $L^+$ vary together experimentally, complicating isolation of their effects. A numerical approach allows independent boundary conditions and systematic variation. We examine the evolution of three-dimensional vortical structures and near-wall behavior across $C_B$, $L^+$, and $D^+$ to inform SJA selection for flow control.

\clearpage

\section{\label{sec:methods}Numerical Methods}

\subsection{\label{sec:governing_equations}Governing equations and turbulence model}

The three-dimensional URANS simulations were conducted using OpenFOAM v2412 \cite{jasak2009}. The unsteady Reynolds-averaged mass and momentum equations are
\begin{equation}
\frac{\partial U_i}{\partial x_i} = 0,
\end{equation}
\begin{equation}
\frac{\partial U_i}{\partial t} + U_j \frac{\partial U_i}{\partial x_j} = -\frac{1}{\rho} \frac{\partial P}{\partial x_i} + \nu \frac{\partial^2 U_i}{\partial x_j \partial x_j} - \frac{\partial}{\partial x_j} \left ( \overline{u'_i u'_j} \right ),
\end{equation}
where $U_i$ are mean velocity components, $\rho$ is density, $P$ is mean pressure, and $\overline{u'_i u'_j}$ are the Reynolds stresses. With the Boussinesq approximation,
\begin{equation}
- \overline{u'_i u'_j} = \nu_t \left( \frac{\partial U_i}{\partial x_j} + \frac{\partial U_j}{\partial x_i} \right) - \frac{2}{3} k\, \delta_{ij},
\end{equation}
\noindent where $\nu_t$ is eddy viscosity, $\delta_{ij}$ the Kronecker delta, and $k$ turbulent kinetic energy. Following preliminary testing in prior work \cite{ho2022,ho2023}, the low-Reynolds-number $k$--$\varepsilon$ model of Launder--Sharma \cite{launder1974} was selected; it models the viscous sublayer via damping functions, avoiding wall functions.

The discretized equations were solved with the \texttt{pisoFoam} solver (finite-volume method). Spatial discretization used second-order schemes for convective terms; first-order upwind was applied to turbulence quantities for stability. Time integration used first-order explicit Euler with adaptive time-stepping to maintain a Courant--Friedrichs--Lewy (CFL) number $<0.9$. A convergence threshold of $10^{-6}$ was applied to all variables.

\subsection{\label{sec:model_setup}Model Setup, Boundary and Test Conditions}

\begin{figure}[ht]
    \includegraphics[width=0.85\textwidth]{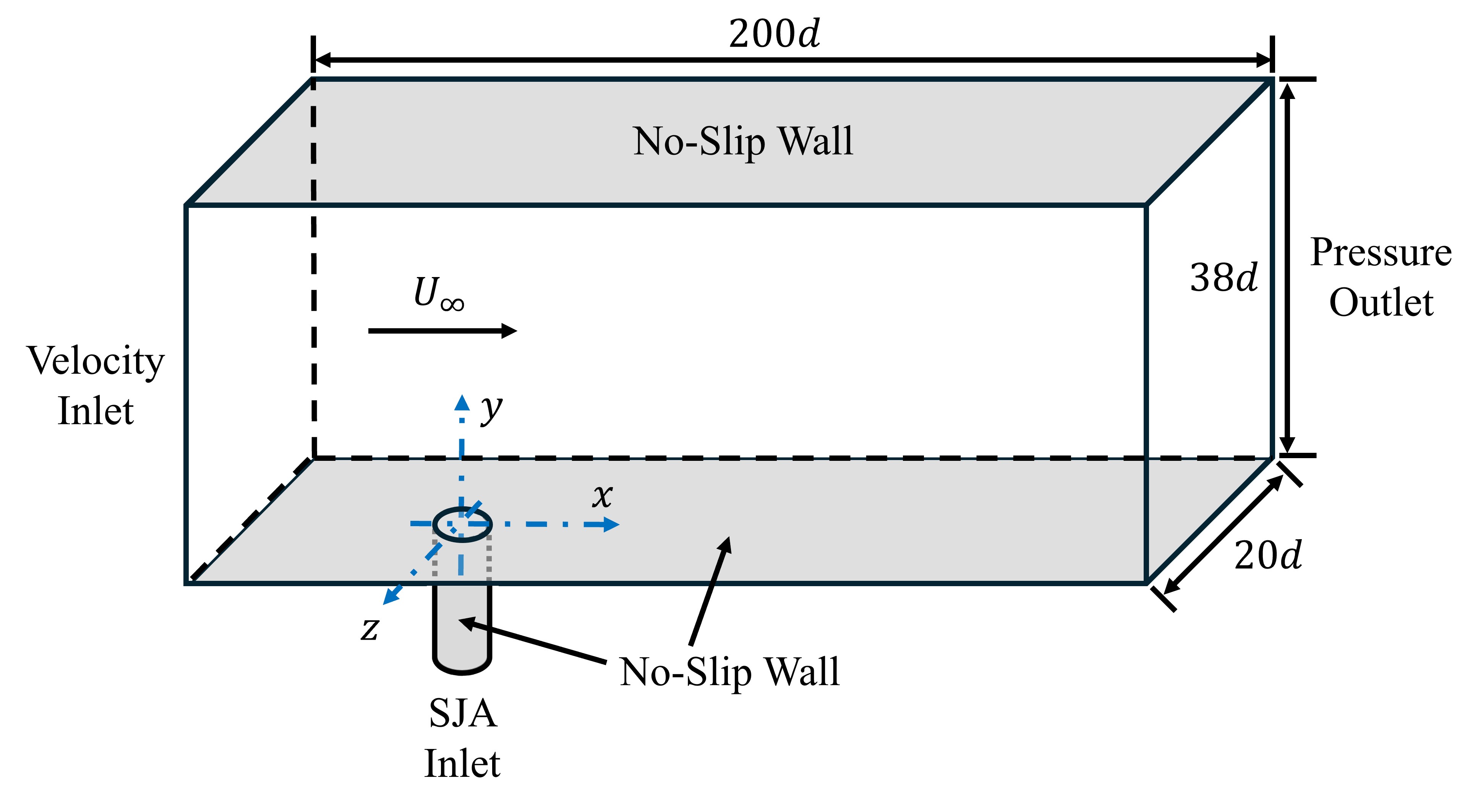}
    \caption{\label{fig:domain}Computational domain and boundary conditions for the SJA in crossflow.}
\end{figure}

The computational domain for the SJA in crossflow is shown in Fig.~\ref{fig:domain}. The rectangular duct is $200d$ long, $20d$ wide, and $38d$ high; the SJA exit center is at $x/d=0$, $y/d=0$, mid-span $z/d=0$, located $25d$ downstream of the inlet. A velocity inlet and pressure outlet were used; walls are no-slip; side boundaries are symmetry. Separate duct simulations generated the different inlet boundary-layer profiles.

The SJA exit diameter is $d = 2~\mathrm{mm}$. The cavity was not modeled; instead, the analytical Womersley solution for pulsating laminar pipe flow was applied at the neck inlet \cite{palumbo2022}:
\begin{equation}
v(r,t) = V_j \, \Re \left\{ 
\left[ 1 - \frac{J_0(i^{3/2}\, W_o \, r)}{J_0(i^{3/2}\, W_o)} \right] e^{i \omega t} 
\right\},
\end{equation}
\noindent where $V_j$ is the maximum centerline velocity, $\omega$ is the angular frequency, $W_o=d\sqrt{\omega/4\nu}$ is the Womersley number, and $J_0$ is the zeroth-order Bessel function. Prior work \cite{ho2023} validated that, with sufficient modeled neck volume, this method reproduces whole-SJA (dynamic mesh) results at lower cost. Here, a neck height ratio $h/d=15$ is used to satisfy the neck-volume requirement and avoid nonphysical jet exit behavior.

\begin{table}[!ht]
\caption{\label{tab:test_cases}Summary of parameters for the synthetic jet in crossflow.}
\begin{ruledtabular}
\begin{tabular}{lcccccccc}
\multicolumn{1}{c}{} & \multicolumn{4}{c}{\textbf{BL parameters}} & \multicolumn{4}{c}{\textbf{SJA parameters}} \\
\cmidrule(lr){2-5} \cmidrule(lr){6-9}
Case & $D^+$ & $Re_\theta$ & $H$ & $Re_\tau$ & $Re_j$ & $C_B$ & $f$ & $L^+$ \\
\hline
A-1 &  &  &  &  & 570  & 0.85 & 200 & 10.6 \\
A-2 & 2.1 & 170 & 3.1 & 49  & 1150 & 1.7  & 200 & 21.3 \\
A-3 &  &  &  &  & 1150 & 1.7  & 400 & 10.6 \\
\hline
B-1 &  &  &  &  & 570  & 0.85 & 200 & 10.6 \\
B-2 & 4.1 & 460 & 1.8 & 135 & 1150 & 1.7  & 200 & 21.3 \\
B-3 &  &  &  &  & 1150 & 1.7  & 400 & 10.6 \\
\hline
C-1 &  &  &  &  & 570  & 0.85 & 200 & 10.6 \\
C-2 & 8.0 & 740 & 1.5 & 266 & 1150 & 1.7  & 200 & 21.3 \\
C-3 &  &  &  &  & 1150 & 1.7  & 400 & 10.6 \\
\end{tabular}
\end{ruledtabular}
\end{table}

Nine cases (Table~\ref{tab:test_cases}) examine the effects of boundary-layer height and SJA settings. The series are grouped by inlet $D^+$: Series A: $D^+=2.1$; Series B: $D^+=4.1$; Series C: $D^+=8.0$. Within each, case “1” is low momentum/low frequency; “2” is high momentum/low frequency; “3” is high momentum/high frequency.

\subsection{\label{sec:validation}Model Validation}

\begin{figure}[!t]
    \includegraphics[width=\textwidth]{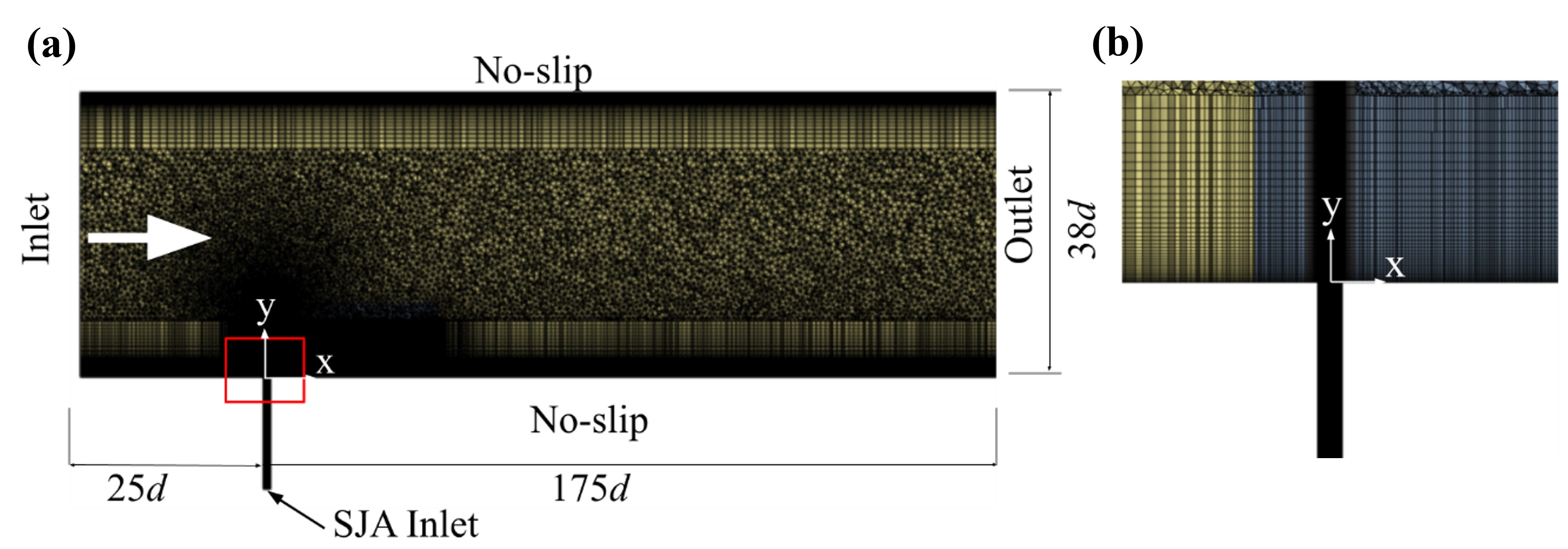}
    \caption{\label{fig:mesh}Mesh configuration: (a) symmetry-plane view (partial length shown) and (b) zoomed view of the jet exit.}
\end{figure}

The computational grid is shown in Fig.~\ref{fig:mesh}. Inflation layers were applied to the top and bottom walls. An O-grid was used in the SJA neck and extended into the duct. Refinement is highest near the exit and immediately downstream, coarsening into the free stream. A grid-resolution analysis \cite{celik2008} on case C-3 used three meshes. Mesh refinement occurred along all axes, focusing on shear-layer and boundary-layer regions. Table~\ref{tab:mesh_study} summarizes centerline velocity $U_{cl}$ at peak expulsion and time-averaged displacement thickness $\delta^*$ at $x/d=10$ for the three meshes. Sampling began after 2 flow-through cycles ($t U_\infty/l = 2$, where $l$ is the length of the duct), which corresponds to a dimensionless time based on jet diameter of $t^*=tU_\infty/d = 400$. Time-averaged $\delta^*$ was computed over five actuation cycles. Grid II was selected for accuracy versus cost. The medium mesh had 40 wall layers, first cell height $75~\mu$m, growth rate 1.07, and a dimensionless first cell wall distance of $y^+_{\max}<5$. 

\begin{table}[ht]
\caption{\label{tab:mesh_study}Grid properties for the mesh-sensitivity study (case C-3).}
\begin{ruledtabular}
\begin{tabular}{lccccc}
Grid & Total cells & $U_{cl}$ & $U_{cl}$ Uncertainty (\%) & $\delta^*_{x/d=10}$ & $\delta^*_{x/d=10}$ Uncertainty (\%) \\
\hline
I   & $2 \times 10^6$  & 15.8 & --   & 4.97 & --   \\
II  & $5 \times 10^6$ & 15.6 & 1.5  & 4.39 & 1.9  \\
III & $13 \times 10^6$ & 15.5 & $<\!1$   & 4.45 & $<\!1$   \\
\end{tabular}
\end{ruledtabular}
\end{table}

\begin{figure}[!t]
    \includegraphics[width=0.65\textwidth]{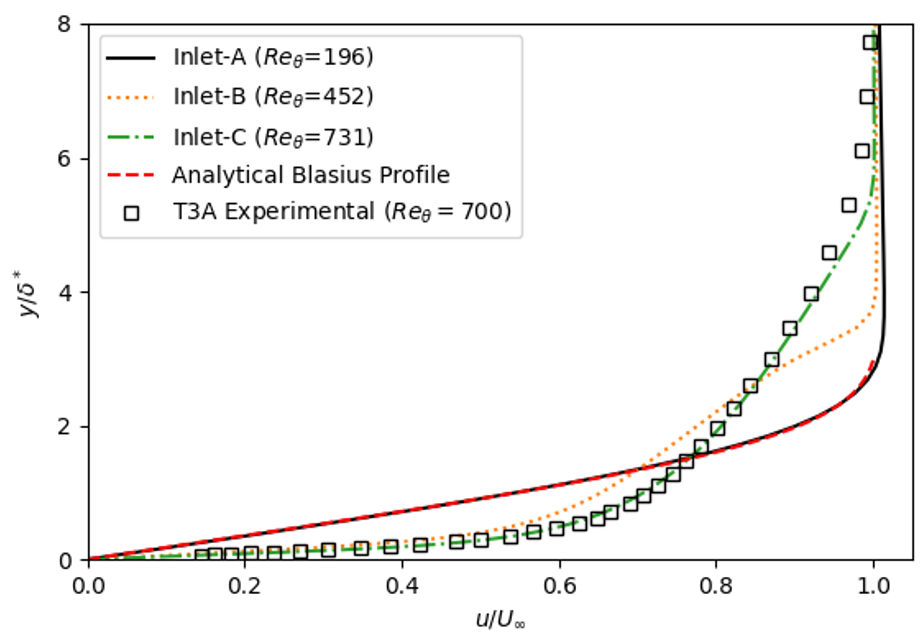}
    \caption{\label{fig:bl_validation}Boundary-layer profiles for the three inlets, compared with analytical/experimental data\cite{roach1992} at similar Reynolds numbers.}
\end{figure}

Baseline boundary layers for the three series of test cases were sampled at $x/d=0$ (Fig.~\ref{fig:bl_validation}). The Blasius profile is used for the lowest-$Re$ inlet A; the T3A experimental data\cite{roach1992} is used to validate the highest-$Re$ inlet C. The agreement between the present simulations and the experimental and analytical results validates the approach boundary-layer.

Sampling of crossflow simulations began after 4.375 flow-through cycles ($t^*= 875$). Probes near the jet exit and within the boundary layer confirmed highly periodic behavior with minimal phase fluctuations; therefore, phase averaging was not performed. Time-averaged fields were obtained by averaging every time step over five actuation cycles.

\clearpage

\section{\label{sec:results}Results and Discussion}

\subsection{\label{sec:flow_structure}Instantaneous Flow Structure}

$Q$-criterion contours are used to visualize vortical structures, where $Q$ is the second invariant of the velocity-gradient tensor~\cite{hunt1988}. Chakraborty et al.~\cite{chakraborty2005} showed that different vortex-identification criteria are qualitatively consistent, validating the use of $Q$ as
\begin{equation}
Q = \tfrac{1}{2}\left( \|\boldsymbol{\Omega}\|^2 - \|\mathbf{S}\|^2 \right) > 0,
\end{equation}
identifying rotation-dominated regions. Here, $Q$ is applied to instantaneous 3-D fields to visualize structures induced by the synthetic-jet boundary-layer interaction.

\begin{figure}[ht]
    \includegraphics[width=\textwidth]{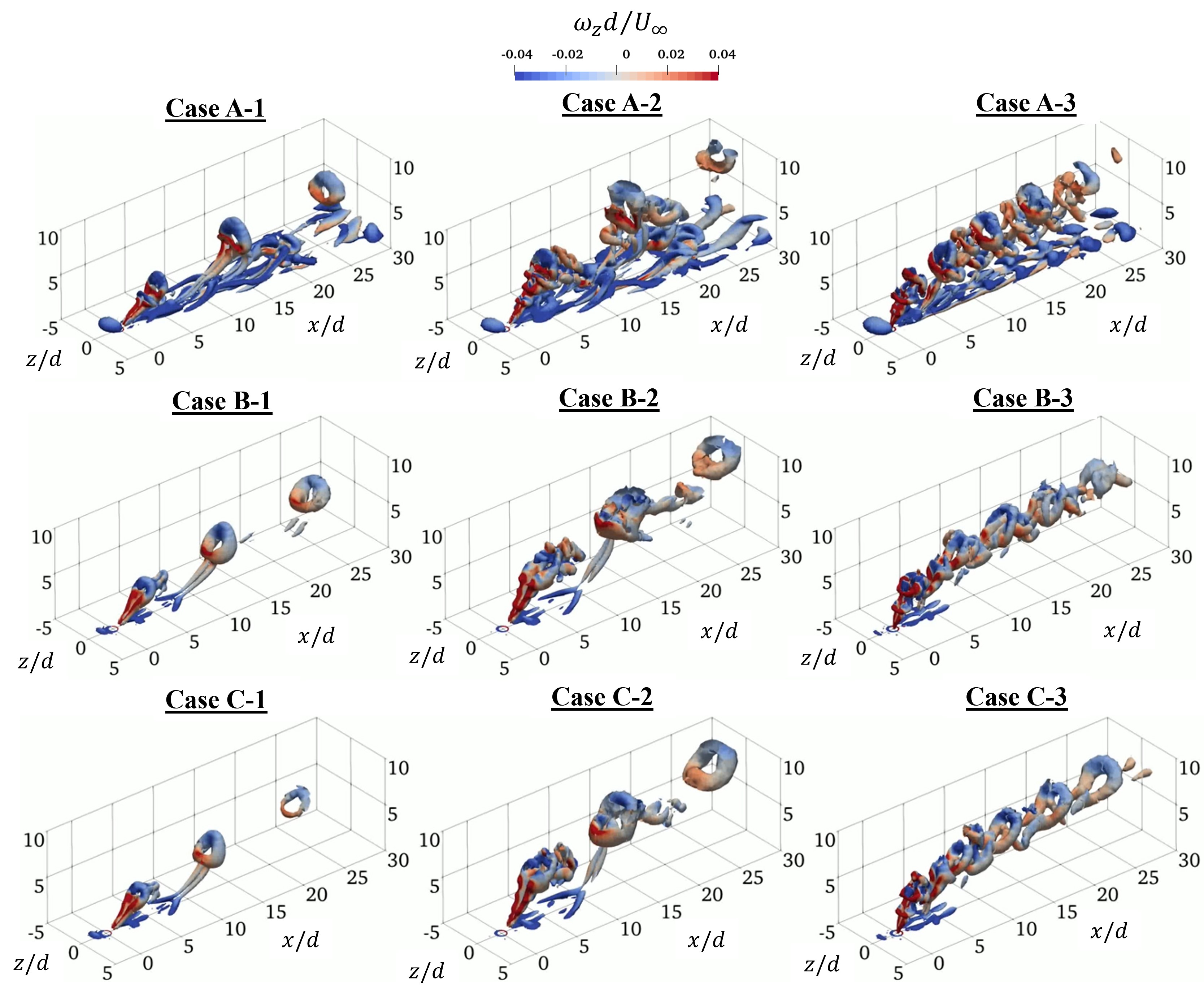}
    \caption{\label{fig:q_cri_iso}Instantaneous $Q$-criterion iso-surfaces ($Q = 0.1 U_\infty^2 / d^2$) colored by spanwise vorticity at isometric view.}
\end{figure}

\begin{figure}[ht]
    \includegraphics[width=\textwidth]{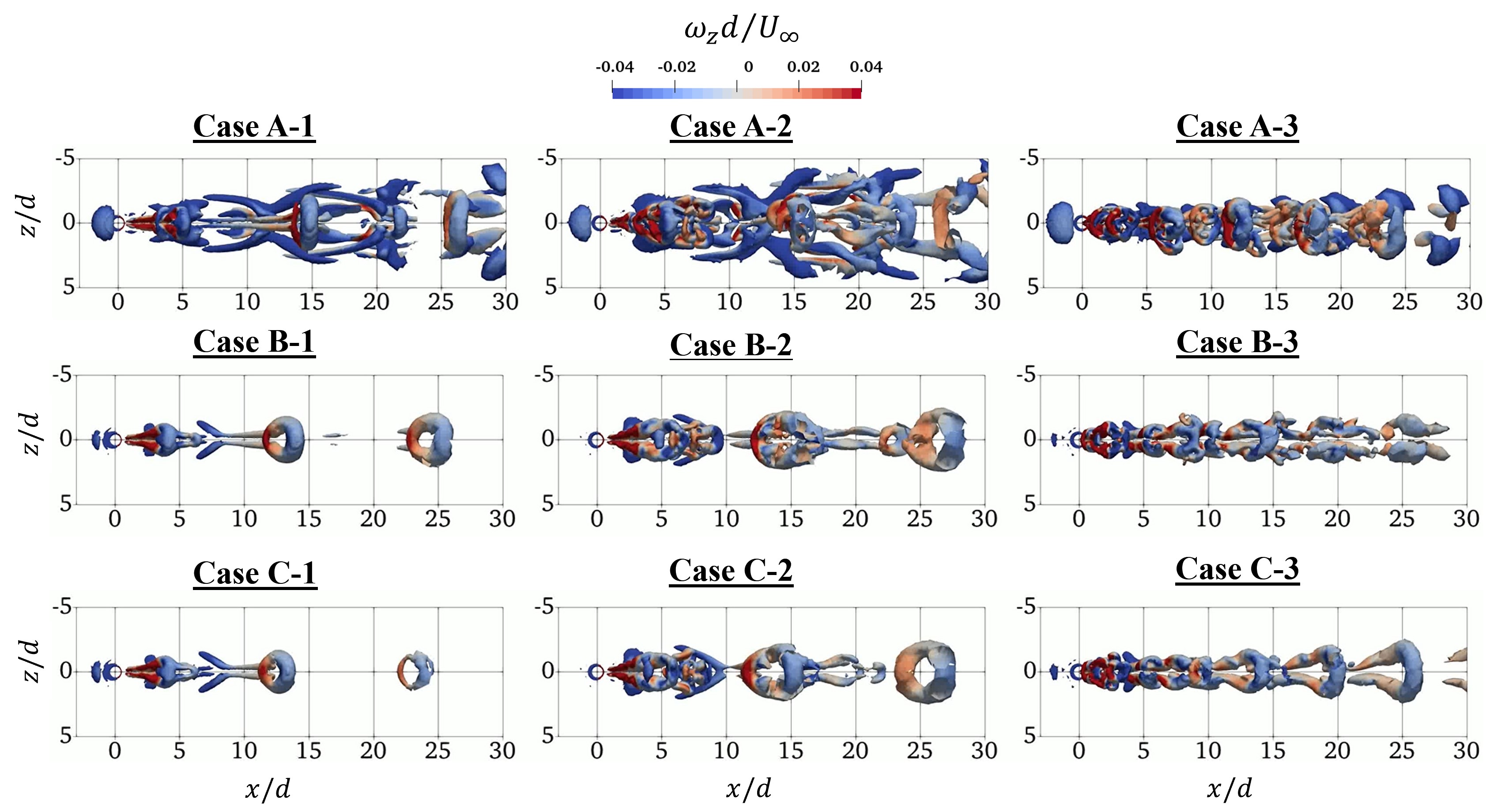}
    \caption{\label{fig:q_cri_top} Instantaneous $Q$-criterion iso-surfaces ($Q = 0.1 U_\infty^2 / d^2$) colored by spanwise vorticity at $x$–$z$ plane view.}
\end{figure}

Figures~\ref{fig:q_cri_iso} and \ref{fig:q_cri_top} show iso-surfaces at $t^*=875$, corresponding to the start of the ingestion cycle for all cases. Low-momentum cases (A-1, B-1, C-1) display a tilted vortex ring (VR) with trailing vortex pairs (TVP) extending toward the wall. With higher momentum (A-2, B-2, C-2), the VR penetrates further; upstream shear-layer interactions produce vortex loops \cite{wu2009,ravi2004}. At high momentum and frequency (A-3, B-3, C-3), expelled structures cluster more closely streamwise. In C-3 (largest $D^+$), VRs break into hairpin-like structures away from the wall due to interference between consecutive structures; this is not observed in A-3 and B-3, where the jet experiences a more uniform crossflow. Top-down views (Fig.~\ref{fig:q_cri_top}) show slower convection of the primary VR with increasing $D^+$ at low momentum; this trend diminishes at higher momentum/frequency.

Near-wall tertiary vortices (TV) increase near-wall momentum by inducing downwash and are relevant to separation control \cite{jabbal2010,ho2022}. In series A, prominent TVs convect downstream along the wall; at higher frequency (A-3), their spanwise footprint narrows, indicating reduced near-wall control authority. With increasing $D^+$ (series B, C), TVs dissipate sooner, consistent with weaker near-wall momentum and reduced shear-layer interaction.

\begin{figure}[p]
    \centering
    \begin{subfigure}{0.85\textwidth}
        \centering
        \includegraphics[width=\linewidth]{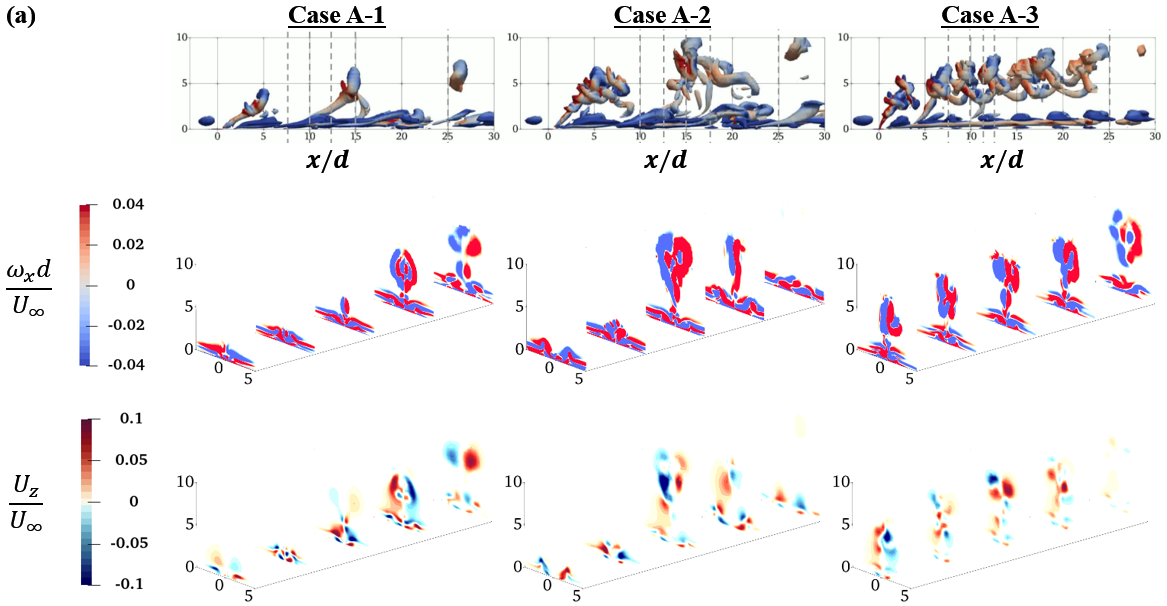}
    \end{subfigure}
    \vspace{0.3em}

    \begin{subfigure}{0.85\textwidth}
        \centering
        \includegraphics[width=\linewidth]{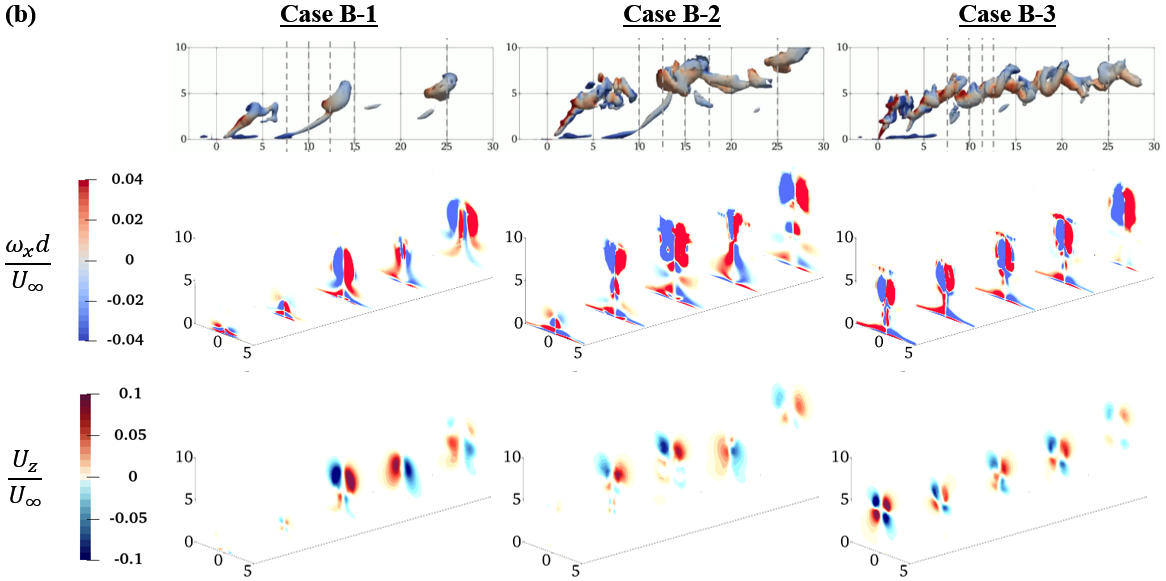}
    \end{subfigure}
    \vspace{0.3em}

    \begin{subfigure}{0.85\textwidth}
        \centering
        \includegraphics[width=\linewidth]{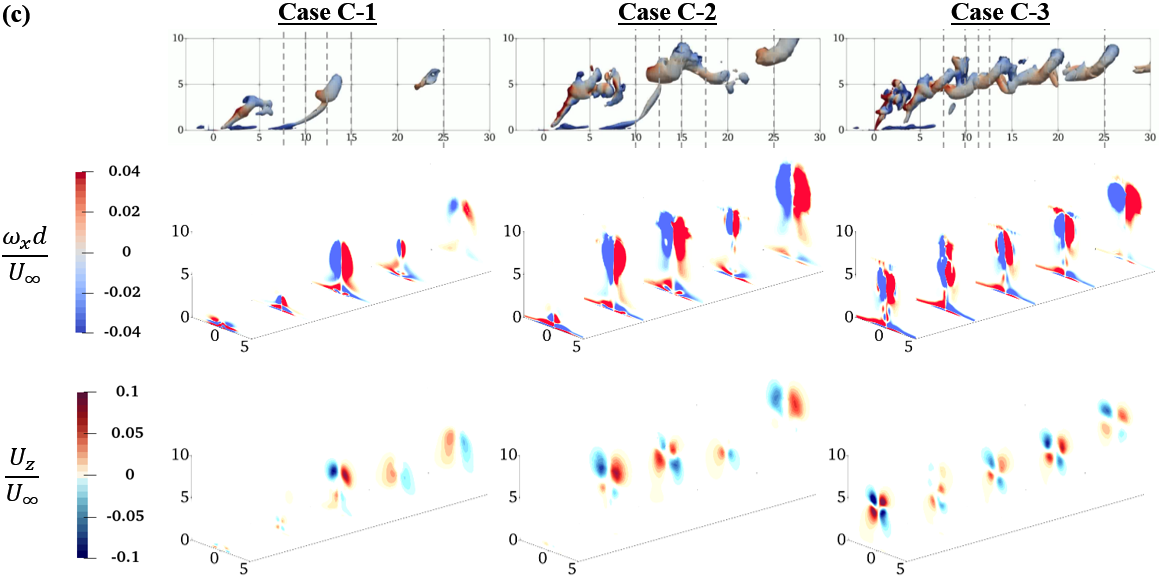}
    \end{subfigure}

    \caption{\label{fig:xy-plane}Instantaneous $Q$-criterion iso-surfaces ($Q = 0.1 U_\infty^2 / d^2$) colored by spanwise vorticity ($x$–$y$ plane); normalized streamwise vorticity and spanwise velocity ($y$–$z$ planes) at indicated $x/d$.}
\end{figure}

Figure~\ref{fig:xy-plane} further probes spanwise control via streamwise vorticity and spanwise velocity. Plane locations (dashed lines) vary with jet momentum/frequency but are fixed with respect to $D^+$ to capture the primary structures. In series A, A-2 shows the deepest penetration, while A-3 exhibits closer-spaced structures and a narrower near-wall spanwise impact than A-1/A-2. In series B and C, near-wall spanwise flow is weaker due to greater $D^+$.

\begin{figure}[ht]
    \includegraphics[width=\textwidth]{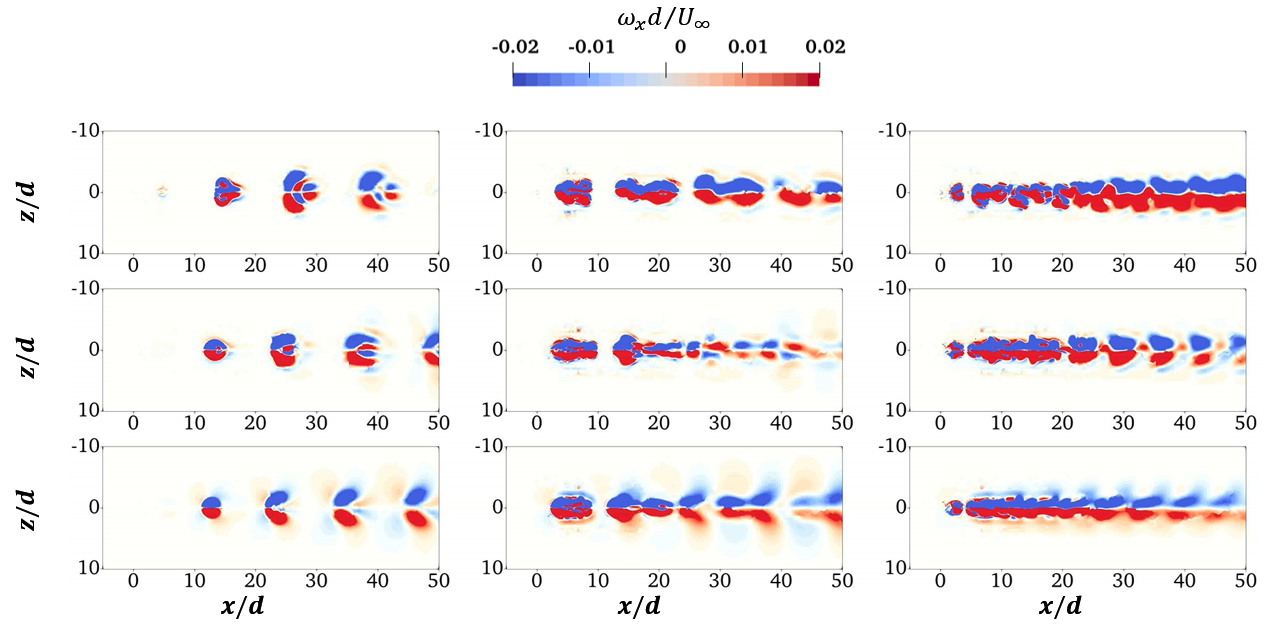}
    \caption{\label{fig:top-down-omega}Instantaneous streamwise vorticity contours at $y/d = 5$.}
\end{figure}

Top-down planes (Fig.~\ref{fig:top-down-omega}) show classic vortex-pair signatures for low-momentum VRs at $y/d=5$. With higher momentum, the vorticity concentrates near mid-span and structures approach continuity at a higher frequency, resembling a steady jet in crossflow.

\subsection{\label{sec:jet_trajectory}Jet Trajectory}

\begin{figure}[ht]
    \includegraphics[width=0.65\textwidth]{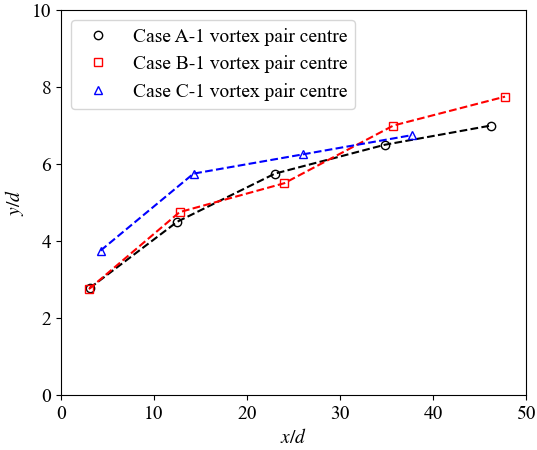}
    \caption{\label{fig:jet_trajectory}Normalized vortex-center trajectories for cases A-1, B-1, and C-1.}
\end{figure}

\begin{figure}[ht]
    \includegraphics[width=\textwidth]{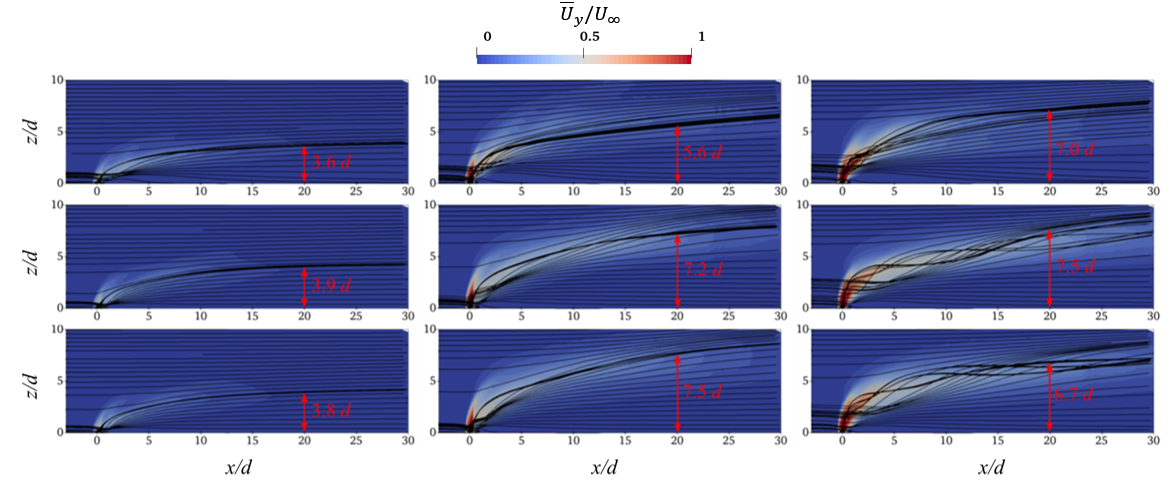}
    \caption{\label{fig:streamlines}Time-averaged normalized transverse velocity along the symmetry plane with time-averaged streamlines superimposed.}
\end{figure}

To examine the effect of $D^+$ on trajectory and penetration, the vortex-pair centers for A-1, B-1, and C-1 are plotted in Fig.~\ref{fig:jet_trajectory}. High-momentum/frequency cases are omitted due to complex symmetry-plane vorticity. The trajectory for C-1 rises more rapidly but plateaus sooner, consistent with Fig.~\ref{fig:q_cri_top}. Time-averaged streamlines superimposed on transverse-velocity contours are shown in Fig.~\ref{fig:streamlines}. The streamlines originating from the jet exit are evaluated up to $x/d=20$. Overall, penetration does not vary strongly with $D^+$, with A-2 an exception showing lower penetration than B-2/C-2.

\subsection{\label{sec:time_averaged}Time-Averaged Boundary-Layer Profile}

\begin{figure}[ht]
    \includegraphics[width=\textwidth]{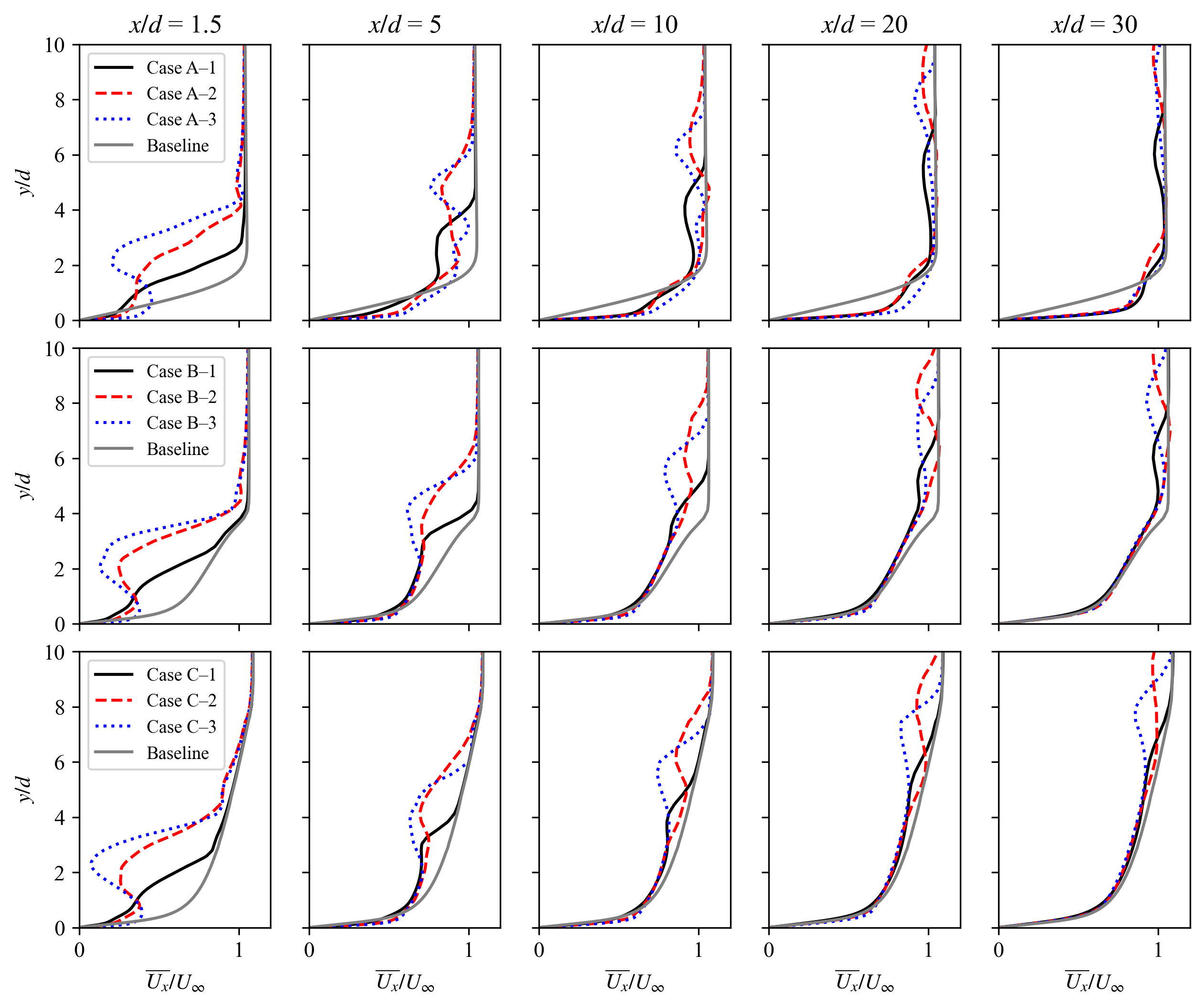}
    \caption{\label{fig:ux-profile}Time-averaged normalized streamwise velocity along the symmetry plane.}
\end{figure}

Figure~\ref{fig:ux-profile} shows time-averaged streamwise velocity $\overline{U}_x/U_\infty$ at several $x/d$. For series A, all SJAs increase near-wall momentum ($y/d\in[0,0.5]$), beneficial for separation delay, with a compensating outer deficit ($y/d\in[0.5,10]$). The outer deficit recovers but remains visible at $x/d=30$; near-wall profiles converge by $x/d=30$. For series B and C, behavior is similar, but the near-wall increase is weaker; by $x/d=30$, the near-wall profiles are close to baseline while an outer deficit remains.

\begin{figure}[ht]
    \includegraphics[width=\textwidth]{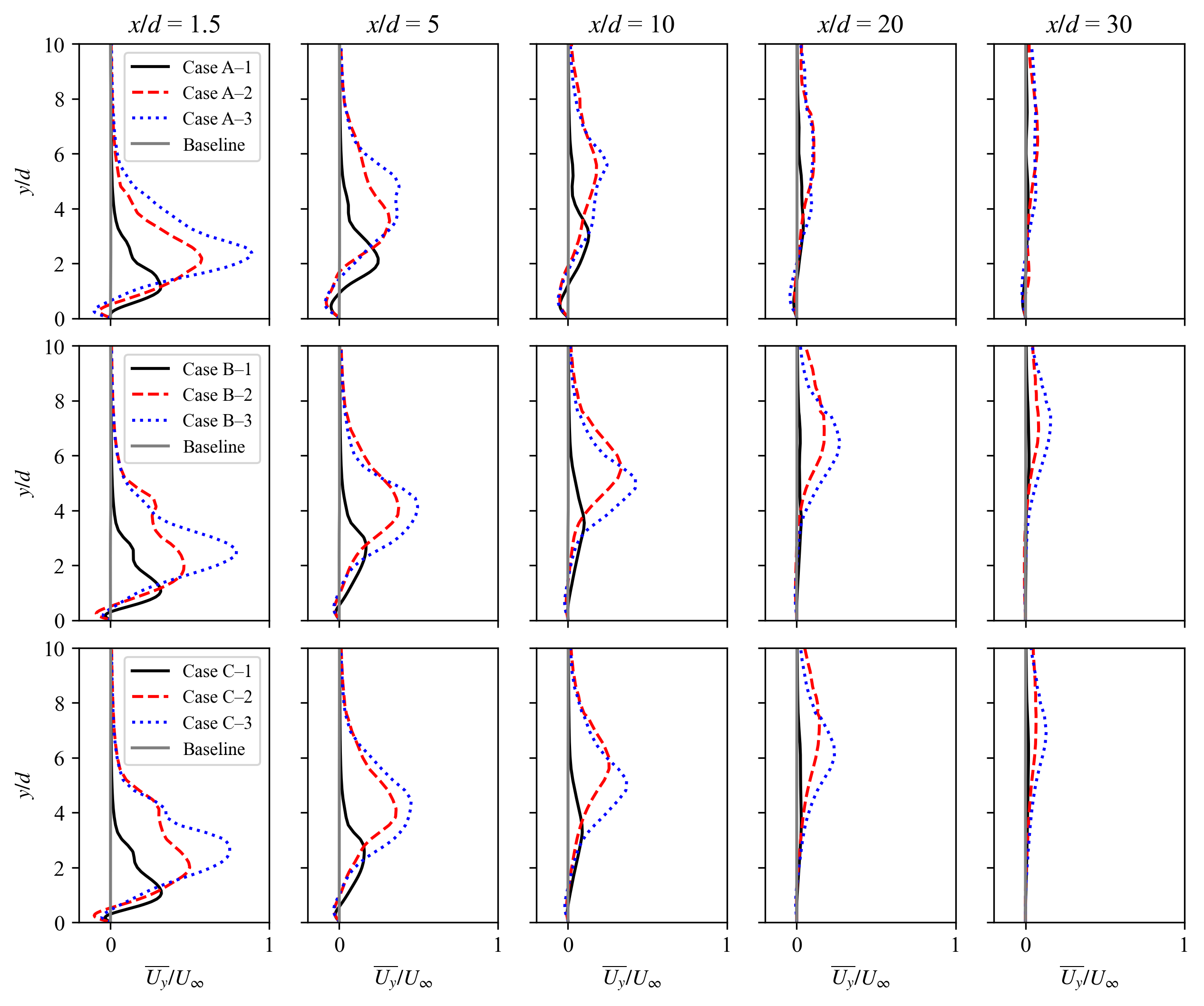}
    \caption{\label{fig:uy-profile}Time-averaged normalized transverse velocity along the symmetry plane.}
\end{figure}

Time-averaged transverse velocity $\overline{U}_y/U_\infty$ is presented in Fig.~\ref{fig:uy-profile}, which shows downwash near the wall ($y/d<1$) at $x/d=1.5$ for all actuated cases, consistent with TVs. In series A, the downwash persists to $x/d=10$; in series B and C, it decays by $x/d=5$. Away from the wall ($y/d>1$), a primary peak develops; a minor secondary peak near the exit for B, C decays by $x/d=5$.

\begin{figure}[ht]
    \includegraphics[width=0.99\textwidth]{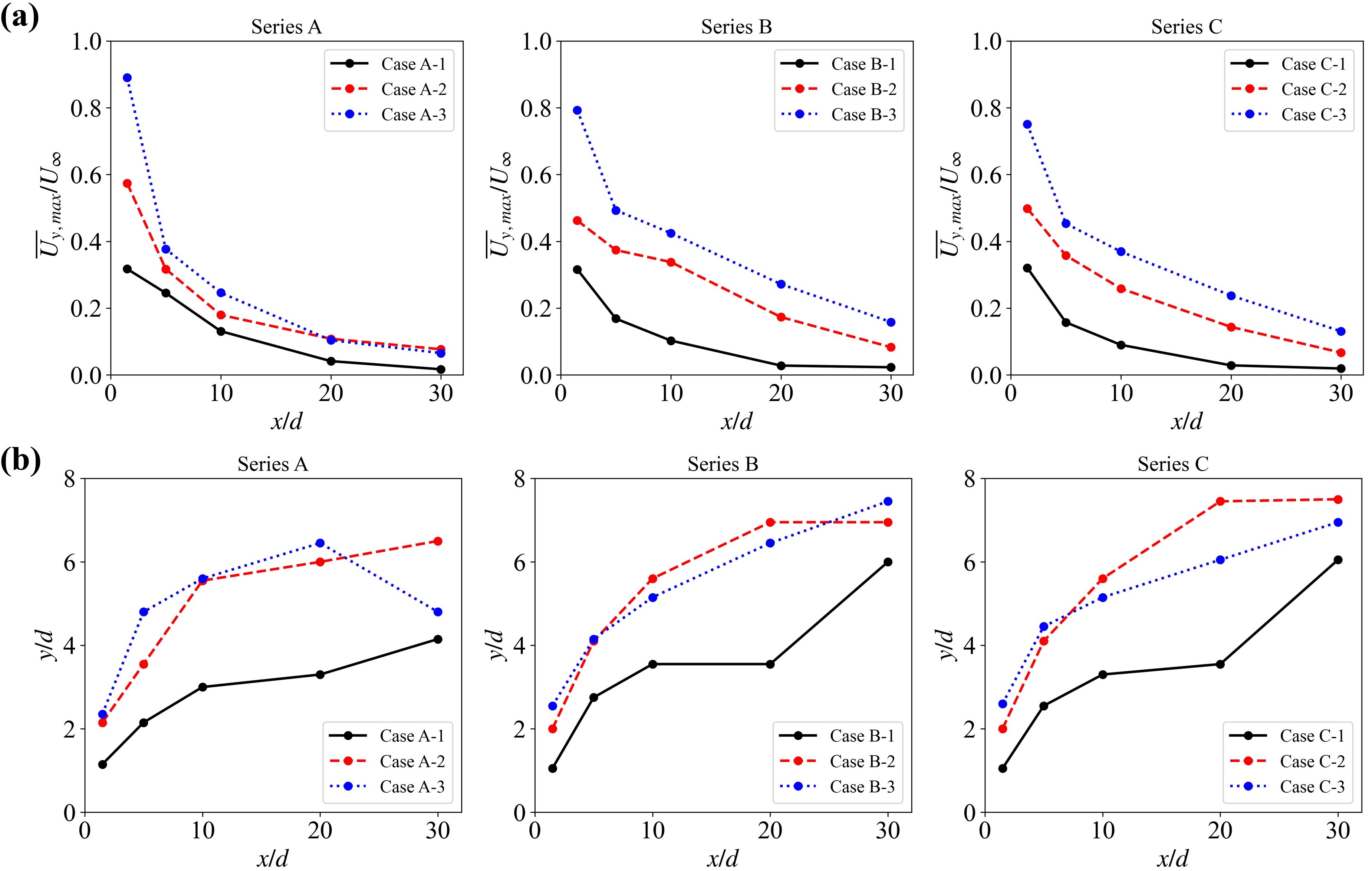}
    \caption{\label{fig:uy-peaks}(a) Downstream decay of $\overline{U}_{y,~\max}/U_\infty$ against \ $x/d$; (b) Wall-normal location $y/d$ of $\overline{U}_{y,~\max}$ against $x/d$.}
\end{figure}

The magnitude and location of the peak transverse velocity are extracted and plotted in Fig.~\ref{fig:uy-peaks}. Peak transverse velocity decays rapidly for all cases from $x/d=1.5$ to $5$, most sharply at high frequency (A-3, B-3, C-3). In series A, the magnitudes of A-2 and A-3 converge after $x/d=20$, but their peak locations differ: for A-3, the peak occurs at $y/d=6.45$ at $x/d=20$ and moves to $y/d=4.8$ by $x/d=30$, whereas A-2 moves farther from the wall with $x$. In series B and C, peak magnitudes and locations vary modestly, except B-3 peaks farther from the wall than B-2 at $x/d=30$. Thus $D^+$ is most influential at low $D^+$ ($\lesssim2$); moderate--high $D^+$ ($\ge 4$) has less effect on transverse profiles.

\begin{figure}[ht]
    \includegraphics[width=1\textwidth]{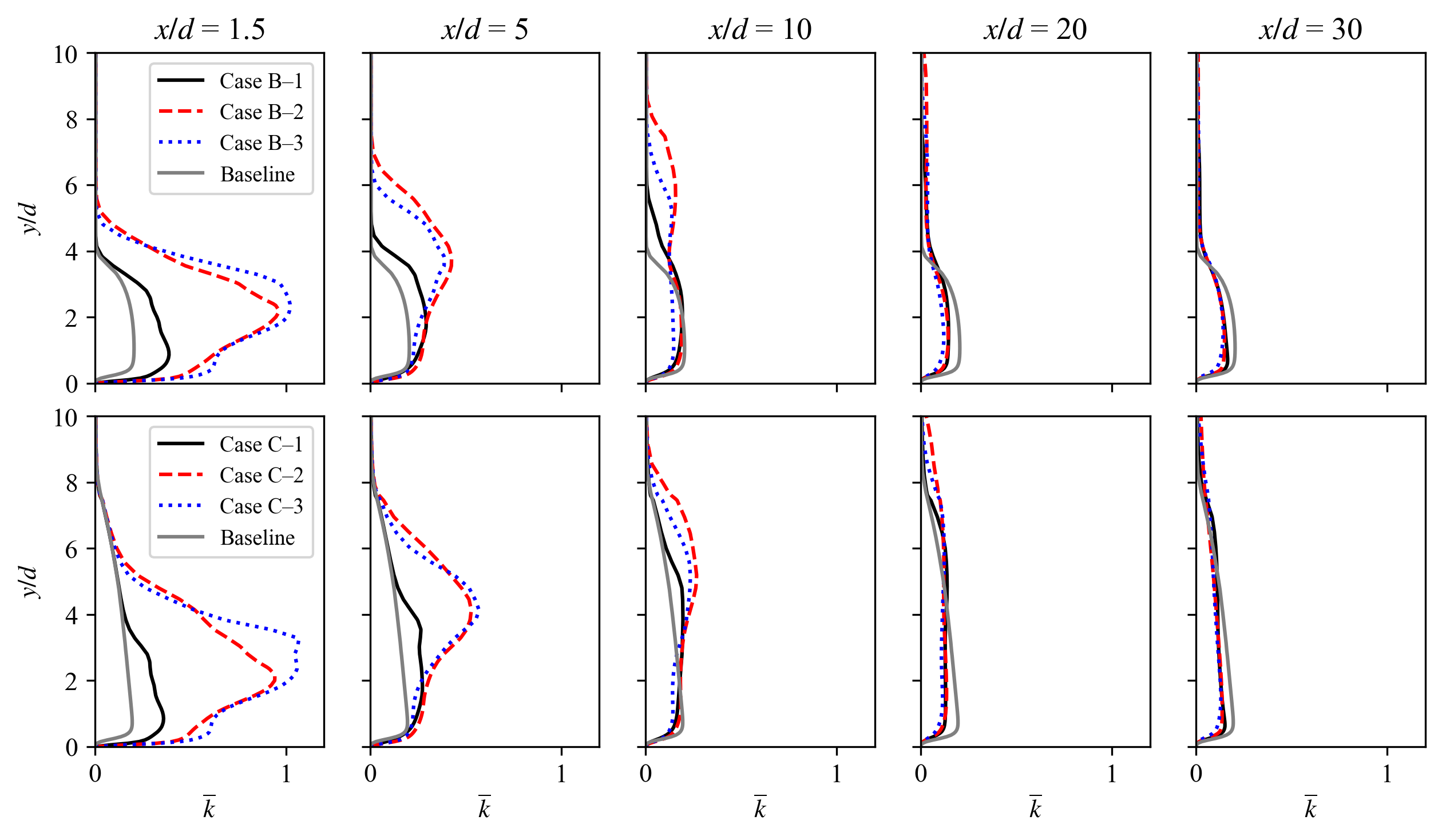}
    \caption{\label{fig:tke-profile} Time-averaged turbulent kinetic energy along the symmetry plane for series B and C.}
\end{figure}

Time-averaged $\overline{k}$ (series B, C) increases in the near field ($x/d\in[1.5,10]$) for all actuated cases, with stronger sensitivity to jet momentum than to $D^+$ or frequency. As structures convect downstream, the SJA effect persists through the boundary layer, with a near-wall $\overline{k}$ deficit appearing for $x/d<20$–$30$.

\clearpage
\subsection{\label{sec:Skin-friction} Skin-Friction Coefficient Profile}

\begin{figure}[ht]
    \includegraphics[width=0.55\textwidth]{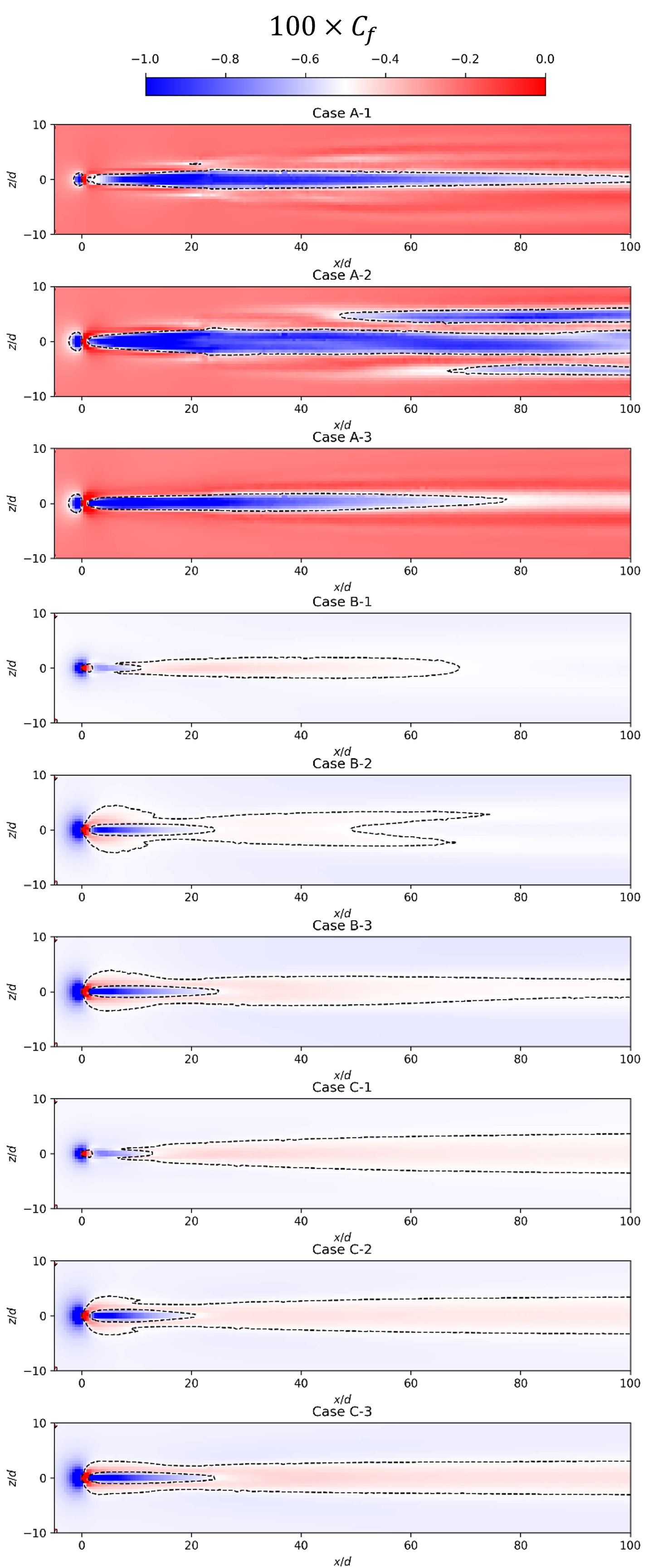}
    \caption{\label{fig:skin_friction}Time-averaged skin-friction coefficient on the lower wall; black dashed contour marks $100\,C_f = -0.5$.}
\end{figure}

Figure~\ref{fig:skin_friction} shows that series A produces large increases in mid-span $C_f$ downstream of the jet exit. Case A-2 shows two additional high-$C_f$ regions at $z/d=\pm5$ not attributable to near-wall hairpins; this suggests excitation of Tollmien--Schlichting (TS) waves. The baseline for series A has $Re_{\delta^*}\approx500$ (critical Blasius $Re_{\delta^*}\approx520$). Following \cite{palumbo2022}, the reduced frequency $F^+=f/f_{\mathrm{ref}}$ with $f_{\mathrm{ref}}=0.752\,U_\infty/d$ gives $F^+\approx1.06$ for A-1/A-2, consistent with TS amplification; the effect is weaker for A-1 due to lower jet momentum. TS amplification is not observed in B-2 or C-2 due to larger $Re_{\delta^*}$.

For B-1 and C-1, $C_f$ increases only within $x/d\lesssim10$; farther downstream, it drops below baseline. Higher momentum (B-2, C-2) extends the region of increased $C_f$ but introduces spanwise deficit regions at moderate--high $C_B$ \cite{ho2022}. Higher frequency (B-3, C-3) yields a modest additional mid-span increase; however, the effect decays rapidly and $C_f$ falls below baseline by $x/d\gtrsim25$, underscoring potential adverse far-field impacts in both spanwise and streamwise directions.

\clearpage

\section{Conclusion}
\label{sec:conclusion}

We independently varied boundary-layer height ratio (\(D^+ \equiv \delta/d\)), blowing ratio (\(C_B\)), and actuation frequency (\(f\)) across nine three-dimensional URANS cases to isolate their individual influence on circular synthetic-jet/crossflow interactions. Three robust trends emerge.

\textit{Actuation frequency governs structure packing and jet penetration.}  
Raising \(f\) from 200\,Hz to 400\,Hz clusters expelled vortex rings in the streamwise direction and reduced jet penetration depth by 15--25\,\% in the thickest layer (\(D^+=8\)). In this regime, consecutive rings interfere and promote a transition from tilted vortex rings to hairpin-like structures farther from the wall. The associated transverse-velocity peaks decay to less than 10\,\% of their near-field magnitude within \(x/d \approx 10\), indicating diminished sustained control authority at high frequency.

\textit{Boundary-layer height modulates near-wall coherence.}  
At low \(D^+ \approx 2.1\), near-wall tertiary vortices persist beyond \(x/d = 20\) and span \(\Delta z/d > 4\), maintaining increased near-wall momentum over long distances. At moderate-to-high \(D^+ \ge 4.1\), these structures dissipate by \(x/d \approx 12\)--15 and near-wall profiles recover to baseline conditions by \(x/d \approx 30\).

\textit{Mean-flow and wall-shear responses reflect a trade-off.}  
Increasing \(C_B\) from 0.85 to 1.7 boosts mid-span skin-friction coefficient (\(C_f\)) by up to 120\,\% immediately downstream (\(x/d < 5\)). However, the spanwise footprint narrows, with off-center deficits emerging at \(|z/d| > 3\). An incidental Tollmien--Schlichting amplification occurs in one low-\(D^+\), low-\(f\) configuration at reduced frequency \(F^+ \approx 1.06\), consistent with transitional receptivity.

\textit{Practical guidance.}  
For flat-plate separation control at \(Re_\theta \sim 10^2\)--\(10^3\), low-to-moderate \(D^+ \le 4\) combined with the lower end of the tested frequency range (\(f \approx 200\)\,Hz) proves most effective for sustained, spanwise-broad near-wall momentum increase. When \(D^+\) is large, raising \(C_B\) helps locally at mid-span but delivers reduced spanwise coverage and faster downstream decay.


\section*{Funding}
This research was funded by the Natural Sciences and Engineering Research Council of Canada (NSERC) grant number RGPIN-2022-03071 and the Digital Research Alliance of Canada (4752).

\begin{acknowledgments}
Computations were performed at the SciNet HPC Consortium.
\end{acknowledgments}

\section*{Data Availability Statement}
The data that support the findings of this study are available from the corresponding author upon reasonable request.

%

\end{document}